\documentclass[aps,prd,groupedaddress,showpacs]{revtex4}

\usepackage{amsfonts}
\usepackage{graphicx}

\begin{document}


\title{Infrared analysis of quantum field theories}


\author{Marco Frasca}
\email[]{marcofrasca@mclink.it}
\affiliation{Via Erasmo Gattamelata, 3 \\ 00176 Roma (Italy)}


\date{\today}

\begin{abstract}
We show that for a $\lambda\phi^4$ theory having many components, the solution
with all equal components in the infrared regime is stable with respect to
our expansion given by a recently devised approach to analyze strongly coupled quantum field theory.
The analysis is extended to a pure Yang-Mills theory showing how, in this case,
the given asymptotic series exists. In this way, many components theories in the
infrared regime can be mapped to a single component scalar field theory obtaining their spectrum.
\end{abstract}


\maketitle


The study of quantum field theories in the infrared regime is generally approached by
pure numerical methods. No perturbative approaches have been known until recently for
an analytical study of these theories. The difficulty can be traced back to the
increasing of the coupling constant with momentum approaching in this case too large values
for perturbation theory to be useful at all.

In a recent work we have proposed a new approach \cite{fra1} that relies on the duality
principle in perturbation theory \cite{fra2,fra3} permitting to use perturbation theory also
when an increasingly large parameter enters into the equations of a field theory. This
approach has been successfully applied in general relativity proving that a gradient
expansion is a strong coupling expansion \cite{fra4}.

This method has permitted to extract the spectrum of a $\lambda\phi^4$ theory 
with a single component in the
infrared regime giving the mass gap in this case. Our aim in this paper is to extend this
approach to multi-components quantum field theories.

So, let us consider the Hamiltonian (here and in the following we take $\hbar=c=1$)
\begin{equation}
    H = \int d^{D-1}x\left[\frac{1}{2}\pi^2+\frac{1}{2}(\partial_x\phi)^2+
	\frac{1}{2}\mu_0^2\phi^2+
	\frac{1}{4}\lambda\phi^4\right].
\end{equation}
After a normalization to $\mu_0$ is taken, Hamilton equations are 
\begin{eqnarray}
\label{eq:phi}
    \partial_t\phi &=& \pi \\ \nonumber
	\partial_t\pi  &=& \partial_x^2\phi -\phi -\lambda\phi^3.
\end{eqnarray}
We take
\begin{eqnarray}
   \tau &=& \sqrt{\lambda}t \\ \nonumber
   \pi &=& \sqrt{\lambda}\left(\pi^{(0)} + \frac{1}{\lambda}\pi^{(1)}+ \frac{1}{\lambda^2}\pi^{(2)} + \ldots\right) \\ \nonumber
   \phi &=& \phi^{(0)} + \frac{1}{\lambda}\phi^{(1)} + \frac{1}{\lambda^2}\phi^{(2)} + \ldots.
\end{eqnarray}
obtaining the set of dual perturbation equations
\begin{eqnarray}
    \partial_{\tau}\phi^{(0)} &=& \pi^{(0)} \\ \nonumber
	\partial_{\tau}\phi^{(1)} &=& \pi^{(1)} \\ \nonumber
    \partial_{\tau}\phi^{(2)} &=& \pi^{(2)} \\ \nonumber
	                 &\vdots&  \\ \nonumber
	\partial_{\tau}\pi^{(0)} &=& -\phi^{(0)3} \\ \nonumber
	\partial_{\tau}\pi^{(1)} &=& -3\phi^{(0)2}\phi^{(1)} + \partial_x^2\phi^{(0)}-\phi^{(0)} \\ \nonumber
	\partial_{\tau}\pi^{(2)} &=& -3\phi^{(0)}\phi^{(1)2}-3\phi^{(0)2}\phi^{(2)} + \partial_x^2\phi^{(1)}-\phi^{(1)}\\ \nonumber
	                 &\vdots& \label{eq:pert}
\end{eqnarray}
and one sees that at the leading order a homogeneous equation rules the dynamics of the field.
Then, one can prove numerically that in the limit $\lambda\rightarrow\infty$ for the classical theory
one can still use a Green function method \cite{fra1} with the leading order homogeneous equation
\begin{equation}
    \ddot G+\lambda G^3=\delta(t),
\end{equation}
having expicitly restated  $\lambda$. This equation has the exact solution
\begin{equation}
\label{eq:gf}
    G(t)=\theta(t)\left(\frac{2}{\lambda}\right)^{\frac{1}{4}}
	{\rm sn}\left[\left(\frac{\lambda}{2}\right)^{\frac{1}{4}}t,i\right].
\end{equation}
and its time reversed version, with $\rm sn$ the snoidal elliptic Jacobi function. 
The quantum field theory is then given by
\begin{equation}
    Z[j]=\exp\left[\frac{i}{2}\int d^Dy_1d^Dy_2\frac{\delta}{\delta j(y_1)}(-\nabla^2+1)\delta^D(y_1-y_2)
    \frac{\delta}{\delta j(y_2)}\right]Z_0[j]
\end{equation}
being
\begin{equation}
    Z_0[j]=\exp\left[\frac{i}{2}\int d^Dx_1d^Dx_2j(x_1)\Delta(x_1-x_2)j(x_2)\right]
\end{equation}
with the Feynman propagator
\begin{equation}
    \Delta(x_2-x_1)=\delta^{D-1}(x_2-x_1)[G(t_2-t_1)+G(t_1-t_2)].
\end{equation}
Noting the following relation for the snoidal function\cite{gr}
\begin{equation}
    {\rm sn}(u,i)=\frac{2\pi}{K(i)}\sum_{n=0}^\infty\frac{(-1)^ne^{-(n+\frac{1}{2})\pi}}{1+e^{-(2n+1)\pi}}
    \sin\left[(2n+1)\frac{\pi u}{2K(i)}\right]
\end{equation}
being $K(i)$ the constant
\begin{equation}
    K(i)=\int_0^{\frac{\pi}{2}}\frac{d\theta}{\sqrt{1+\sin^2\theta}}\approx 1.3111028777,
\end{equation}
the Fourier transform of the Feynman propagator is
\begin{equation}
\label{eq:prop}
    \Delta(\omega)=\sum_{n=0}^\infty\frac{B_n}{\omega^2-\omega_n^2+i\epsilon}
\end{equation}
being
\begin{equation}
    B_n=(2n+1)\frac{\pi^2}{K^2(i)}\frac{(-1)^{n+1}e^{-(n+\frac{1}{2})\pi}}{1+e^{-(2n+1)\pi}},
\end{equation}
and the mass spectrum of the theory given by
\begin{equation}
\label{eq:ms}
    \omega_n = \left(n+\frac{1}{2}\right)\frac{\pi}{K(i)}\left(\frac{\lambda}{2}\right)^{\frac{1}{4}}
\end{equation}
proper to a harmonic oscillator. In this case, the mass gap in the limit $\lambda\rightarrow\infty$ is
\begin{equation}
\label{eq:ds}
     \delta_S = \frac{\pi}{2K(i)}\left(\frac{\lambda}{2}\right)^{\frac{1}{4}}
\end{equation}
corresponding to the choice $n=0$, that is produced by the self-interaction of the scalar field. As shown
in Ref.\cite{fra1}, the next correction in the dual perturbation series is proportional to the
missing term in the propagator, i.e. ${\bf k}^2+1$ as it should be.

We now go one step further with respect to \cite{fra1} proving that the perturbation series
given in eqs.(\ref{eq:pert}) is meaningful by solving them. This series should prove to give
increasingly smaller terms with increasing $\lambda$. The existence of a classical solution
in the strong coupling limit is essential to prove that, already at this stage, no
instabilities set in and also to look at the analytical form of this solution. In this way
we can be sure that also a quantum field theory in the same limit can be built. Our aim
is to prove existence and we do not care about explicit Lorentz invariance but we note that
all this computations can be carried out with explicit Lorentz invariance by introducing
a fifth variable and working with an Euclidean theory\footnote{I have to thank
Hagen Kleinert for pointing me out this.}.

Then, we can write immediately the solution at the leading order by imposing Lorentz invariance as
\begin{equation}
   \phi^{(0)}=A{\rm sn}\left(\frac{A}{\sqrt{2}}(\sqrt{\lambda}t-{\bf k}\cdot {\bf x}),i\right)
\end{equation}
that is a Jacobi snoidal wave with a wave vector ${\bf k}$ and a normalization factor dependent on the energy as
also happens for ordinary plane waves. Normalization on a box will yield a quantization of the wave
vector. The next order will give the equation to be solved
\begin{equation}
    \partial^2_\tau\phi^{(1)}+3\phi^{(0)2}\phi^{(1)}=-{\bf k}^2\phi^{(0)3}-\phi^{(0)}.
\end{equation}
What we have to prove is that this term is bounded so that the strong coupling expansion is meaningful in the
classical model. For our aims we will content with a WKB solution to the above equation. This is consinstent
with the requirement of a perturbation going to infinity (formally $\lambda\rightarrow\infty$)\cite{fra2}.
This gives
\begin{equation}
    \phi^{(1)}\approx -\sqrt{\frac{2}{3}}\frac{1}{A^2}
    \tan\left(\sqrt{3}\int_0^\tau d\tau'\phi^{(0)}(\tau',{\bf x})\right)
    \frac{{\bf k}^2\phi^{(0)}(\tau,{\bf x})^3+\phi^{(0)}(\tau,{\bf x})}
    {{\rm cn}\left(\frac{A}{\sqrt{2}}(\tau-{\bf k}\cdot {\bf x}),i\right)
    {\rm dn}\left(\frac{A}{\sqrt{2}}(\tau-{\bf k}\cdot {\bf x}),i\right)}.
\end{equation}
We can see that the solution is finite except for those sets of points where WKB approximation
is expected to fail, that is, caustics. Then, the next correction can be written down straightforwardly as
\begin{equation}
    \phi^{(2)}\approx \sqrt{\frac{2}{3}}\frac{1}{A^2}
    \tan\left(\sqrt{3}\int_0^\tau d\tau'\phi^{(0)}(\tau',{\bf x})\right)
    \frac{\partial^2_x\phi^{(1)}(\tau,{\bf x})-\phi^{(1)}(\tau,{\bf x})-3\phi^{(0)}(\tau,{\bf x})\phi^{(1)}(\tau,{\bf x})^2}
    {{\rm cn}\left(\frac{A}{\sqrt{2}}(\tau-{\bf k}\cdot {\bf x}),i\right)
    {\rm dn}\left(\frac{A}{\sqrt{2}}(\tau-{\bf k}\cdot {\bf x}),i\right)}.
\end{equation}
that behaves similarly to the first correction. The next orders can be written down in a similar
manner. The conclusion is that we have built a strong coupling solution of the classical
equation of a scalar field with a single component. The series exists, apart for caustics,
and is meaningful. Then, whenever we will be able to map a field theory to this
case, a meaningful quantum field theory in the infrared is obtained and the spectrum
in this regime can be immediately derived.

In order to exploit this approach, let us consider a U(1)-invariant scalar field theory. This
theory has the Hamiltonian
\begin{equation}
  H = \int d^{D-1}x\left[\frac{1}{2}\pi_1^2+\frac{1}{2}\pi_2^2
  +\frac{1}{2}(\partial_x\phi_1)^2+\frac{1}{2}(\partial_x\phi_2)^2
  +\frac{1}{2}\mu_0^2(\phi_1^2+\phi_2^2)
  +\frac{1}{4}\lambda(\phi_1^2+\phi_2^2)^2\right].
\end{equation}
This gives the following Hamilton equations (after normalization to $\mu_0$)
\begin{eqnarray}
    \partial_t\phi_1 &=& \pi_1 \\ \nonumber
	\partial_t\pi_1  &=& \partial_x^2\phi_1 -\phi_1 -\lambda\phi_1(\phi_1^2+\phi_2^2) \\ \nonumber
    \partial_t\phi_2 &=& \pi_2 \\ \nonumber
	\partial_t\pi_2  &=& \partial_x^2\phi_2 -\phi_2 -\lambda\phi_2(\phi_1^2+\phi_2^2). \label{eq:phi2}
\end{eqnarray}
Applying the above procedure we get the perturbation equations
\begin{eqnarray}
	\partial_{\tau}^2\phi_1^{(0)} &=& -\phi_1^{(0)}(\phi_1^{(0)2}+\phi_2^{(0)2}) \\ \nonumber
	\partial_{\tau}^2\phi_1^{(1)} &=& -(3\phi_1^{(0)2}+\phi_2^{(0)2})\phi_1^{(1)} 
	-2\phi_1^{(0)}\phi_2^{(0)}\phi_2^{(1)} 
	+\partial_x^2\phi_1^{(0)}-\phi_1^{(0)} \\ \nonumber
	\partial_{\tau}^2\phi_1^{(2)} &=&  
	-3\phi_1^{(0)2}\phi_1^{(2)}-3\phi_1^{(0)}\phi_1^{(1)2}
	-2\phi_1^{(1)}\phi_2^{(0)}\phi_2^{(1)}-2\phi_1^{(0)}\phi_2^{(0)}\phi_2^{(2)}
	-\phi_1^{(0)}\phi_2^{(1)2}-\phi_1^{(2)}\phi_2^{(0)2}
	+ \partial_x^2\phi_1^{(1)}-\phi_1^{(1)}\\ \nonumber
	                 &\vdots& \\ \nonumber
	\partial_{\tau}^2\phi_2^{(0)} &=& -\phi_2^{(0)}(\phi_1^{(0)2}+\phi_2^{(0)2}) \\ \nonumber
	\partial_{\tau}^2\phi_2^{(1)} &=& -(3\phi_1^{(0)2}+\phi_2^{(0)2})\phi_2^{(1)} 
	-2\phi_1^{(0)}\phi_2^{(0)}\phi_1^{(1)} 
	+\partial_x^2\phi_2^{(0)}-\phi_2^{(0)} \\ \nonumber
	\partial_{\tau}^2\phi_2^{(2)} &=&   
	-3\phi_2^{(0)2}\phi_2^{(2)}-3\phi_2^{(0)}\phi_2^{(1)2}
	-2\phi_2^{(1)}\phi_1^{(0)}\phi_1^{(1)}-2\phi_1^{(0)}\phi_2^{(0)}\phi_1^{(2)}
	-\phi_2^{(0)}\phi_1^{(1)2}-\phi_2^{(2)}\phi_1^{(0)2} 
	+ \partial_x^2\phi_2^{(1)}-\phi_2^{(1)}\\ \nonumber
	                 &\vdots& \label{eq:pert2}
\end{eqnarray}
So, a fully mapping of this perturbation series with the one of a single scalar field
can be obtained by looking for a solution with all the components being equal. This in turn
means that the spectrum in the infrared of this theory is obtained by taking into eq.(\ref{eq:ms})
the substitution $\lambda\rightarrow 2\lambda$. This is a general rule as for a 
theory with $N$ components one has to take the substitution  $\lambda\rightarrow N\lambda$
into the spectrum giving the result we were looking for. This result holds because both
the classical and the quantum field theory for a single component scalar field do exist as we
have shown and we can map a N-component scalar field theory on it. We note as the presence
of a natural energy scale $\mu_0$ makes the spectrum of the theory not dependent on any
arbitrary parameter. Things will be different for a Yang-Mills theory as we will see below.

We now apply the above procedure to a Yang-Mills theory for the gauge group SU(N). We note
that in this case the only constant entering into the equation, the charge, can be removed
from the equations. Anyhow, we will keep it as a placeholder for the series terms. Our
computation can be just carried out when the gauge is fixed. So,
the Hamilton equations in the gauge $A_0^a=0$ can be written down as \cite{fs,smi}
\begin{eqnarray}
    \partial_t A_k^a&=&F_{0k}^a \\ \nonumber
    \partial_t F_{0k}^a&=&\partial_lF_{lk}^a+gf^{abc}A_l^bF_{lk}^c
\end{eqnarray}
being $g$ the coupling constant, $f^{abc}$ the structure constants of the gauge group,
$F_{lk}^a=\partial_lA_k^a-\partial_kA_l^a+gf^{abc}A_l^bA_k^c$ and the constraint
$\partial_kF_{0k}^a+gf^{abc}A_k^bF_{0k}^c=0$ does hold. So, let us introduce the
following equations, as done for the scalar field,
\begin{eqnarray}
   \tau &=& gt \\ \nonumber
   F_{0k}^a&=& gF_{0k}^{a(0)} + F_{0k}^{a(1)} + \frac{1}{g}F_{0k}^{a(2)} + \ldots \\ \nonumber
   F_{lk}^a&=& F_{lk}^{a(0)} + \frac{1}{g}F_{lk}^{a(1)} + \frac{1}{g^2}F_{lk}^{a(2)} + \ldots \\ \nonumber
   A_k^a &=& A_k^{a(0)} + \frac{1}{g}A_k^{a(1)} + \frac{1}{g^2}A_k^{a(2)} + \ldots.
\end{eqnarray}
Then, one has the perturbation equations
\begin{eqnarray}
    \partial_\tau A_k^{a(0)}&=&F_{0k}^{a(0)} \\ \nonumber
    \partial_\tau A_k^{a(1)}&=&F_{0k}^{a(1)} \\ \nonumber
    &\vdots& \\ \nonumber
    \partial_\tau F_{0k}^{a(0)}&=&f^{abc}f^{cde}A_l^{b(0)}A_l^{d(0)}A_k^{e(0)} \\ \nonumber
    \partial_\tau F_{0k}^{a(1)}&=&
    f^{abc}f^{cde}A_l^{b(1)}A_l^{d(0)}A_k^{e(0)}
    +f^{abc}f^{cde}A_l^{b(0)}A_l^{d(1)}A_k^{e(0)}
    +f^{abc}f^{cde}A_l^{b(0)}A_l^{d(0)}A_k^{e(1)} \\ \nonumber
    & &+f^{abc}\partial_l\left(A_l^{b(0)}A_k^{c(0)}\right)
    +f^{abc}A_l^{b(0)}\left(\partial_lA_k^{c(0)}-\partial_kA_l^{c(0)}\right) \\ \nonumber
    &\vdots&
\end{eqnarray}
and we can recognize at the leading order the homogeneuos Yang-Mills equations. These equations
display a rich dynamics as e.g. Hamiltonian chaos \cite{sav1,sav2,sav3}. It is commonly
accepted that when equal components are considered as a leading order solution, this solution
is unstable\footnote{I was not able to trace this claim in the literature.}. 
Here we have proved that this is not true with respect to our series. Indeed, we are
able to map Yang-Mills theory on a single scalar field theory by taking all equal components
and the series exists both as a classical and as a quantum field theory as we have already seen.
But the mapping can be just completed by the introduction of an arbitrary energy scale $\Lambda$
that can be only obtained through experiment or by a lattice computation. Here we derive its
value through the mass of the lightest unflavored resonance. In order to obtain the spectrum we easily realize
that the right substitution is given by $\lambda\rightarrow g^2N\Lambda^4$. This gives the
spectrum
\begin{equation}
\label{eq:dm}
     \omega_n = \left(n+\frac{1}{2}\right)\frac{\pi}{K(i)}\left(\frac{g^2N}{2}\right)^{\frac{1}{4}}\Lambda.
\end{equation}
We see here the proper 't Hooft scaling $Ng^2$
as also shown in lattice Yang-Mills quantum field theory computations \cite{tep}. 
This gives all the spectrum of a 0$^{++}$ glueball. We just note that the ground state $n=0$ gives the
mass scale of the propagator \cite{cuc}.

So, we give the glueball
spectrum to be compared with \cite{tep2}. We get from eq.(\ref{eq:dm}) for the scalar
glueball and its excited states
\begin{equation}
    \frac{m_G}{\sqrt{\sigma}}=1.198140235(2n+1)
\end{equation}
where we have put the tension $\sigma=\Lambda^2\sqrt{g^2N/2}$.

So, let us compare the ground state $n=0$ with the lightest unflavored resonance $f_0(600)$. As recently
shown \cite{cola} one should have in this case $M_\sigma=441^{+16}_{-8}$ MeV with same
magnitude order as given in \cite{morn}. This gives $\sqrt{\sigma}\approx 368$ MeV, 
in agreement with \cite{cuc}, that can be used to compute the mass of the $0^{++}$ glueball.

Indeed, other excited states are given by $n=1$, $n=2$ and $n=3$ as
\begin{eqnarray}
    \frac{m_G}{\sqrt{\sigma}}(n=1)&=&3.594420705 \\ \nonumber
	\frac{m_G}{\sqrt{\sigma}}(n=2)&=&5.990701175 \\ \nonumber
	\frac{m_G}{\sqrt{\sigma}}(n=3)&=&8.386981645
\end{eqnarray} 
in very good agreement with the values in ref.\cite{tep2} $3.55(7)$ and $5.69(10)$ for
SU(3) and $3.307(53)$ and $6.07(17)$ for the extrapolated values for $N$ to infinity. Finally,
the $0^{++}$ mass is given for $n=1$ by $m_G=1323$ MeV.

A note about the gluon propagator in the infrared is needed. Indeed, we can see that this is
finite and does not go to zero (\ref{eq:prop}). Rather, it is in nice agreement with recent analysis \cite{bou}. 

In conclusion we have proved a general approach to analyze multi-component quantum field theories in
infrared regime obtaining their spectrum with respect to an effective single component
scalar quantum field theory. When applied to a pure Yang-Mills theory the agreement with lattice
computations turns out to be very satisfactory. It would be really interesting to see a solution
of a scalar quantum field theory on a lattice to study its infrared properties in the light of our
results.


\end{document}